# Magnetic Nanoparticle Relaxation Dynamics-based Magnetic Particle Spectroscopy (MPS) for Rapid and Wash-free Molecular Sensing

Kai Wu[†, *], Jinming Liu[†], Diqing Su[‡], Renata Saha[†], and Jian-Ping Wang[†, *]

[†]Department of Electrical and Computer Engineering, University of Minnesota, Minneapolis, Minnesota 55455, USA

[‡]Department of Chemical Engineering and Material Science, University of Minnesota, Minneapolis, Minnesota 55455, USA

**Abstract**

Magnetic nanoparticles (MNPs) have been extensively used as contrasts and tracers for bioimaging, heating sources for tumor therapy, carriers for controlled drug delivery, and labels for magnetic immunoassays. Here, we describe a MNP Brownian relaxation dynamics-based magnetic particle spectroscopy (MPS) method for the quantitative detection of molecular biomarkers. In MPS measurements, the harmonics of oscillating MNPs are recorded and used as a metric for the freedom of rotational motion, which indicates the bound states of the MNPs. These harmonics can be collected from microgram quantities of iron oxide nanoparticles within 10 seconds. Since the harmonics are largely dependent on the quantity of the MNPs in the sample, the MPS bioassay results could be biased by the deviations of MNP quantities in each sample, especially for the very low concentration biomarker detection scenarios. Herein, we report three MNP concentration/quantity-independent metrics for characterizing the bound states of MNPs in MPS. Using a streptavidin-biotin binding system as a model, we demonstrate the feasibility of using MPS and MNP concentration/quantity-independent metrics to sense these molecular interactions, showing this method can achieve rapid, wash-free bioassays, and is suitable for future point-of-care (POC), sensitive, and versatile diagnosis.

1. Introduction

In recent years, magnetic nanoparticles (MNPs) have been successfully applied as nano-heaters for hyperthermia therapy, nano-carriers for drug delivery, nano-tracers for magnetic particle imaging (MPI), nano-contrast agents for magnetic resonance imaging (MRI), and nano-labels for magnetic bioassays.[1–27] MNPs, with physical size comparable to biologically important substances, have many unique physicochemical properties such as high surface to volume ratio and size-dependent magnetic properties, making them ideal for many novel applications. Nowadays, MNPs with a properly functionalized surface can be physically and chemically stable, biocompatible, and environmentally-friendly.[28] Furthermore, biological samples exhibit virtually no magnetic background, thus high sensitivity measurements can be performed on minimally processed samples in MNP-based biomedical applications. In addition, the ease of synthesis and facile surface chemistry have generated much eagerness in applying MNPs to clinical diagnostics and therapy.

Magnetic particle spectroscopy (MPS) is a novel measurement method that closely relates to MPI, which has been widely explored by many groups in recent years.[29–41] In MPS, sinusoidal magnetic fields periodically drive MNPs into magnetically saturated regions which generate magnetic responses that contain not only the drive field frequencies but also a series of harmonic frequencies. These harmonic components can be easily extracted by means of filtering and fast Fourier transform (FFT). The harmonics are very useful metrics of the MNP ferrofluids which contain important information such as the viscosity and temperature of MNP solution as well as the conjugation of any ligands/chemicals onto MNPs.[30,42–50] It has been reported that since the harmonic amplitude is largely dependent on the quantity of the MNPs in the sample, the MPS bioassay results could be biased by the deviations of MNP quantities in each sample, especially for the very low concentration biomarker detection scenarios.[30,30,34] Due to this concern, it has been reported that magnetic susceptibility is MNP quantity-independent for MNP-based immunoassays.[51–53] Herein, we are reporting three MNP concentration/quantity-independent metrics for characterizing the bound states of MNPs in MPS, namely, the $f_{H,crit}$ at peak harmonic amplitude (unit: Hz), ½FWHM (unit: Hz), and the ratios of the 3$^{rd}$ to the 5$^{th}$ harmonics (R35). We used a MNP Brownian relaxation dynamics-based MPS method as the model for the rapid and wash-free molecular sensing. This type of measurement relies on the fact that the tendency of MNPs' magnetic moments to align with the external magnetic field is countered by the Brownian relaxation that randomizes the MNPs' alignment. The extent of disorder caused by Brownian relaxation is most sensitive to the bound state of MNPs, namely, the hydrodynamic size. The higher harmonics (the 3$^{rd}$ and 5$^{th}$ harmonics) are extracted from the MPS for analyzing the bound states of MNPs.

We verified the feasibility of using these MNP concentration/quantity-independent metrics in MNP Brownian relaxation-based MPS molecular sensing. The well-characterized streptavidin and biotin system with ultra-high binding affinity is chosen as the model system, demonstrating the validity of MPS and MNP concentration/quantity-independent metrics for future rapid, point-of-care (POC), sensitive, and versatile immunoassays. The results show that MNP Brownian relaxation-based MPS method can detect analytes directly from biological samples with minimum sample preparation in a wash-free manner.

2. **Materials and Methods**

**Material.** The MNPs used in this work are iron oxide nanoparticles (IONPs) with average core size of 30 nm and coated with a layer of biotin, purchased from Ocean NanoTech, San Diego, California (catalog no. SHB-30). The streptavidin from *Streptomyces avidinii* is a salt-free, lyophilized powder with biotin binding capacity of 13 units/mg protein, purchased from Sigma-Aldrich inc., Atlanta, Georgia (product no. S4762). Phosphate buffered saline (PBS) is purchased from Sigma-Aldrich inc., Atlanta, Georgia (product no. 79378).

**Sample Preparation.** The lyophilized streptavidin powder is reconstituted in PBS and prepared with different concentrations vary from 75 nM to 15 μM. As shown in Scheme 1, six samples (vial I - VI) are prepared and each sample contains approximately 100 μL MNPs. Vials I - V are active groups and each vial is added with 100 μL streptavidin of different concentrations. Vial VI serves as control group and it is added with 100 μL PBS. All the samples are mixed well and incubated at room temperature for 30 minutes to allow the binding of streptavidin to biotins from MNPs.

Scheme 1. Composition of six MNP samples.

| Vial # | I | II | III | IV | V | VI |
|---|---|---|---|---|---|---|
| Streptavidin Concentration | 15 μM | 7.5 μM | 3.8 μM | 1.5 μM | 75 nM | 0 (PBS) |

**Conjugation of Streptavidin to MNPs.** Streptavidin is a crystalline tetrameric protein, with a molecular weight of 4 × 15 kDa, it binds four molecules of biotin and has a high binding affinity to biotin ligands (dissociation constant $K_d = 10^{-14}\ M$). The binding between biotin and streptavidin is very fast, and once formed, it is independent of temperature, solvents, pH, and other denaturing agents. Each streptavidin hosts 4 biotin binding sites, which allows the interaction with multiple biotin moieties from different MNPs and, as a result, forms MNP clusters.

**Experimental Setups.** The MPS measurement system setups and signal chain are shown in Figure 1(a) & (b). A personal computer (PC) controls the data acquisition card (DAQ, NI USB-6289) to generate two sinusoidal signals, which are amplified by two instrument amplifiers (IA, HP 6824A), followed by two band pass filters (BPFs) to suppress higher harmonics that might be introduced by IAs. These amplified and filtered sinusoidal signals drive the outer and inner coils (see Figure 1(c) & (d)) to generate oscillating magnetic fields: one with frequency $f_L = 10\ Hz$ and amplitude $A_L = 170\ Oe$, the other with frequency $f_H$ varies from 500 Hz to 20 kHz and amplitude $A_H = 17\ Oe$. One pair of differentially wound pick-up coils (600 windings in clock-wise and 600 windings in counter-clock-wise) collect the induced voltage and phase signals from MNPs and send back to a BPF before digitalized on the DAQ. The response signals at combinatorial frequencies $f_H \pm 2f_L$ (3rd harmonic) and $f_H \pm 4f_L$ (5th harmonic) are analyzed.

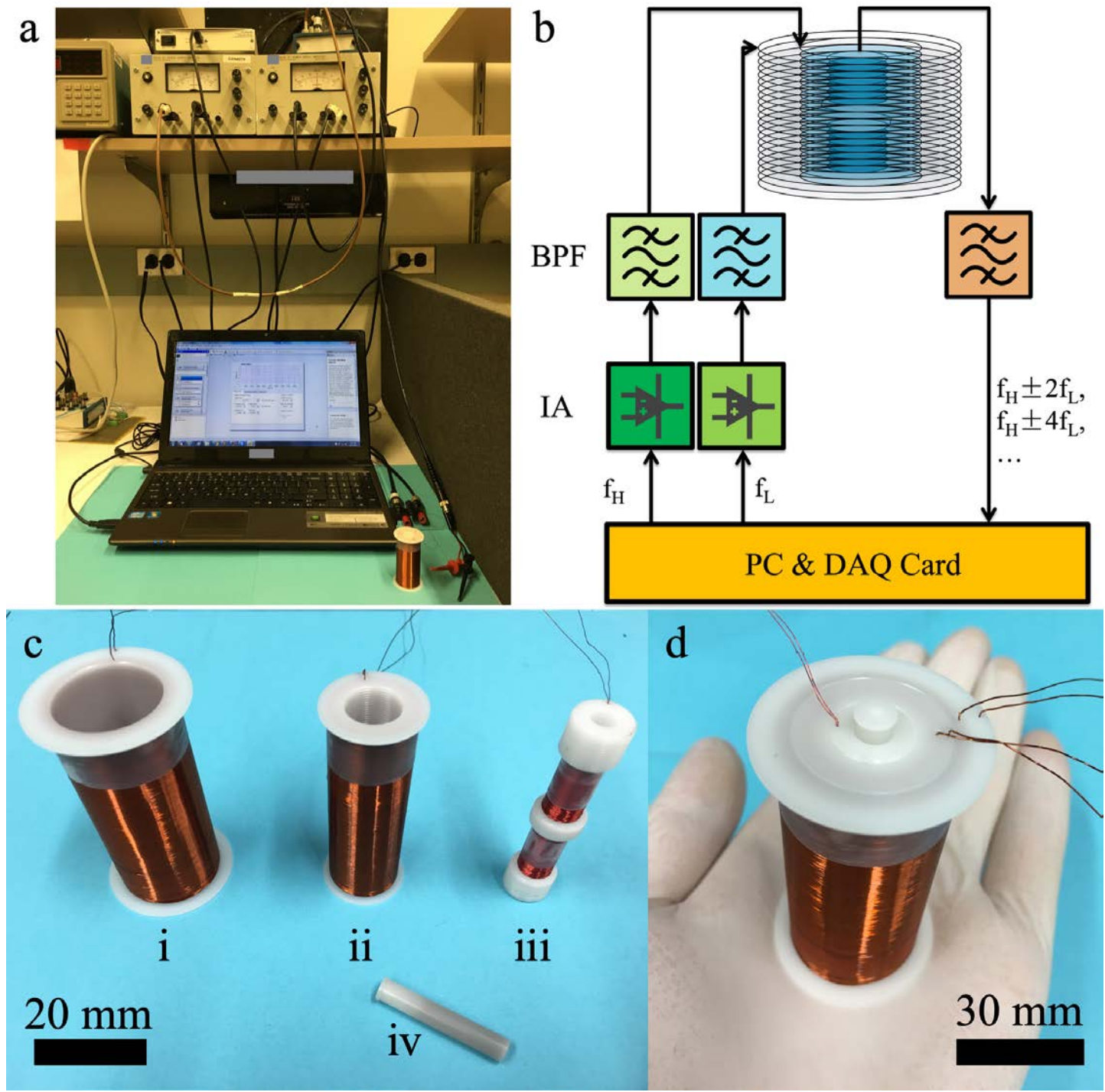

Figure 1. MPS measurement system setups. (a) Photograph of system setups. (b) Signal chain. The system used in this work records nonlinear magnetic response of MNPs under two oscillating magnetic fields with frequencies

of $f_H$ and $f_L$. (c) Photograph of coils: (i) low frequency drive coil (outer coil); (ii) high frequency drive coil (inner coil); (iii) a pair of pick-up coils; (iv) plastic vial with a capacity of 300 μL. (d) Photograph of assembled coils.

**Molecular Sensing via MPS.** When the single domain MNPs are suspended in solution and subjected to an external magnetic field, there are two mechanisms by which the magnetic moments rotate in response to the magnetic field (see Scheme 2(a) & (b)): the intrinsic Néel motion (rotating magnetic moment inside a stationary particle) and the extrinsic Brownian motion (rotating the entire particle along with its magnetic moment). In principle, both the Néel and Brownian mechanism play a role in determining the magnetization of MNP ferrofluids subjected to external oscillating magnetic fields.[32] In this work, we use MNPs with core size of 30 nm, where the Brownian mechanism is dominant and the Néel mechanism is minimized. For these Brownian-relaxation-dominated MNPs, their MPS responses are most sensitive to the hydrodynamic sizes of MNPs. As shown in Scheme 2(c) & (d), when the target biomarker (i.e., streptavidin) has multiple binding sites, ligands (i.e., biotins) from more than one MNP can bind to the same biomarker, resulting in the clustering of MNPs. Such an interaction greatly increases the hydrodynamic sizes of MNPs as well as the Brownian relaxation time. As a result, noticeable changes could be found from the MPS responses due to this specific binding process. Mathematical models of the relaxation mechanisms can be found in Notes S1 & S2 in the Supporting Information.

Scheme 2. (a) Néel relaxation is the rotation of magnetic moment inside a stationary MNP. (b) Brownian relaxation is the rotation of entire MNP along with the magnetic moment. (c) The streptavidin has a high binding affinity to the biotin ligands on MNP surface. (d) As the quantity of streptavidin increases in the MNP suspension, the MNPs are likely to form clusters. The dashed lines represent the hydrodynamic sizes of MNPs due to the clustering induced by streptavidin. As the MNP clustering level increases, the hydrodynamic size increases, and the harmonic amplitude decreases.

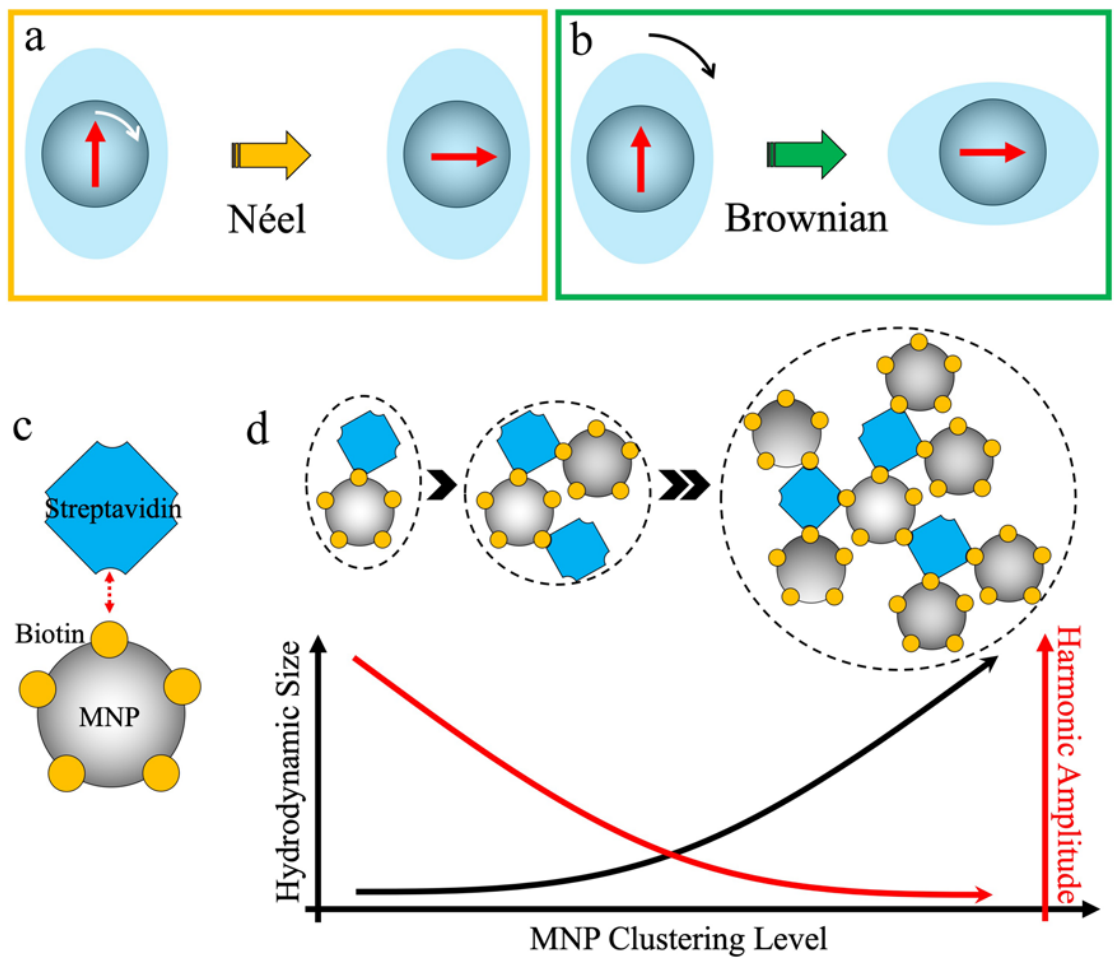

## 3. Results and Discussions

**MPS Characterization of Magnetic Relaxation Dynamics.** The MPS response upon increasing streptavidin concentrations from 75 nM to 15 μM is investigated, and one control group is added in this experiment to verify this detection strategy: biotin coated MNPs with the addition of PBS. The frequency of high frequency drive field $f_H$ is varied at 500 Hz, 1 kHz, 2 kHz, 4 kHz, 6 kHz, 8 kHz, 10 kHz, 12 kHz, 14 kHz, 16 kHz, 18 kHz, and 20 kHz, and the frequency of the low frequency drive field $f_L$ is set at 10 Hz. The amplitudes of both high and low frequency drive fields are identical in each test. During each test, the background noise is monitored for 10 s, then the plastic vial containing 200 μL sample is inserted into the pick-up coils and followed by another 10 s of data collection on the total signal. Some examples of the real-time magnetic responses sensed by pick-up coils are shown in Note S8 in the Supporting Information. The MPS response of MNPs is extracted by subtracting the background noise from the total signal using the phasor theory we reported before:[37,54,55]

$$A_{MNP}e^{j\varphi_{MNP}}\big|_m = A_{TOT}e^{j\varphi_{TOT}}\big|_m - A_{Noise}e^{j\varphi_{Noise}}\big|_m \quad (1),$$

Where $A_{MNP}$, $A_{TOT}$, $A_{Noise}$ are the amplitudes from MNPs, total signal, and the background noise, respectively, and $\varphi_{MNP}$, $\varphi_{TOT}$, and $\varphi_{Noise}$ are the phase angles from MNPs, total signal, and the background noise, respectively, $m$ represents the m[th] measured harmonic. Phasor model can be found in Note S4 in the Supporting Information.

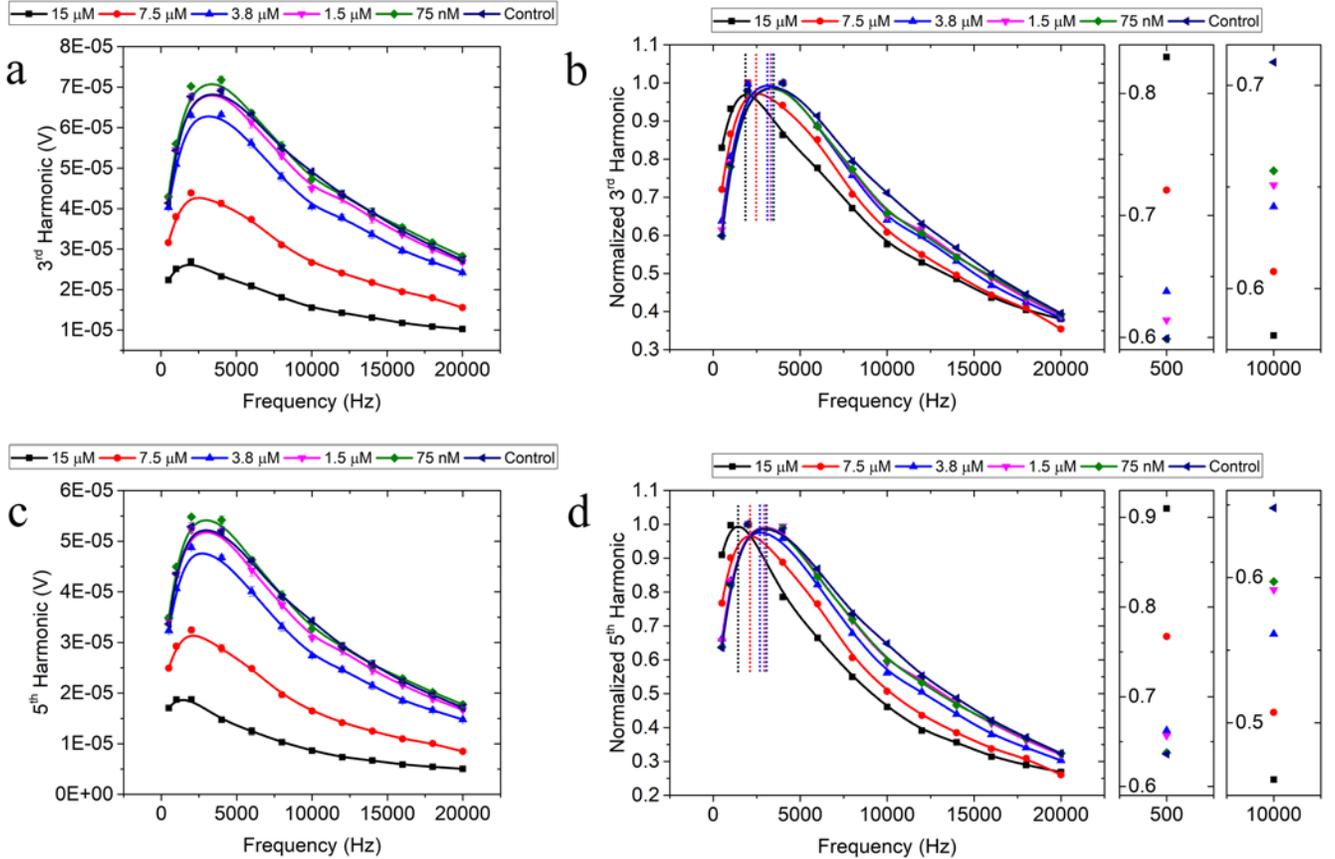

Figure 2. (a, c) MPS measurements of the 3[rd] and 5[th] harmonics from six MNP samples in different bound states. Error bar represents standard deviation. (b, d) MPS measurements of the normalized 3[rd] and 5[th] harmonics from six MNP samples in different bound states. The insets in (b) & (d) highlight the normalized harmonic amplitudes measured at 500 Hz and 10 kHz, respectively. Dotted line represents the position of peak harmonic signal for each sample as the drive field frequency varies.

The 3[rd] and 5[th] harmonic amplitudes of MNPs from six samples are reconstructed from the total signals and background noise, they are summarized in Figure 2(a) & (c). According to the Debye model and Faraday's law of induction (mathematical models of the MPS responses can be found in Notes S3 - S5 in the Supporting Information), the harmonic amplitude is dependent on the drive field frequency $f_H$, the cosine of phase lag $\varphi$, and the quantity of MNPs in the testing sample. The increased streptavidin concentration/quantity in the MNP sample causes larger MNP clusters, thus, the hydrodynamic size of MNPs increases and the phase lag increases, which, as a result, causes noticeable drop in harmonic amplitudes (see Scheme 2(d)). As shown in Figure 2(a) & (c), the harmonic amplitudes of the 3$^{rd}$ and 5$^{th}$ harmonics decreases as we increase the concentration/quantity of streptavidin in the MNP sample. However, the harmonic amplitudes from vial V (75 nM) are larger than those of the control sample. Which is due to the deviations of MNP quantities in each sample. Since the harmonic amplitude is largely dependent on the quantity of the MNPs from the sample, the testing results could be biased by the deviations of MNP quantities in each sample, especially for the very low concentration biomarker detection scenarios. Thus, in this paper, we used the MNP quantity-independent metrics: the normalized 3$^{rd}$ and 5$^{th}$ harmonics and the ratios of the 3$^{rd}$ to the 5$^{th}$ harmonics (R35) to quantify the concentrations of target biomarker.

Firstly, we report here the normalized 3$^{rd}$ and 5$^{th}$ harmonics as a MNP quantity-independent metric for biomarker detection, which are plotted in Figure 2(b) & (d). The dotted lines represent the critical frequencies $f_{H,crit}$ where the harmonic amplitudes reach to peaks for six MNP samples. There is a clear trend that the peaks move to lower $f_H$ for samples with higher concentrations/quantities of streptavidin (explained in Note S9 in the Supporting Information). As is summarized in Table I, the $f_{H,crit}$ at peak harmonic amplitude decreases as the biomarker concentrations/quantities increase.

The insets in Figure 2(b) & (d) highlight the normalized 3$^{rd}$ and 5$^{th}$ harmonic amplitudes measured at 500 Hz and 10 kHz, respectively. We compared the normalized amplitude values from six MNP samples before and after the onset of the peaks. It is found that before the onset of peaks (at 500 Hz), the measured normalized harmonic amplitudes arranged from highest to lowest are: I (15 μM) > II (7.5 μM) > III (3.8 μM) > IV (1.5 μM) > V (75 nM) > VI (control, 0 nM). However, after the onset of peaks (at 10 kHz), the measured normalized harmonic amplitudes reversed, namely, from highest to lowest are: VI (control, 0 nM) > V (75 nM) > IV (1.5 μM) > III (3.8 μM) > II (7.5 μM) > I (15 μM), which indicates that the peak widths from the normalized 3$^{rd}$ and 5$^{th}$ harmonic curves vary for six MNP samples in different bound states. In view of this, we introduce the full width at half maximum (FWHM) as another MNP quantity-independent metric for characterizing the biomarker concentrations/quantities from MNP samples. Since the normalized harmonic amplitude curves in Figure 2(b) & (d) are not full pulse waveforms, we use ½FWHM, which is the difference between $f_{H,crit}$ at peak harmonic amplitude and $f_{H,50\%}$ at 50% of the peak harmonic amplitude. It is observed from Figure 2(b) & (d) that the ½FWHM decreases as the concentrations/quantities of streptavidin increases in the MNP sample. In other words, the normalized harmonic amplitude curve is steeper from MNP samples with higher concentrations/quantities of biomarkers. As is summarized in Table I, the ½FWHM value increases as the biomarker concentrations/quantities increases.

As per best knowledge, it is for the first time that $f_{H,crit}$ at peak harmonic amplitude and ½FWHM as MNP quantity-independent metrics are being used for characterizing the biomarker concentrations/quantities.

Table I. Summarized peak shift of harmonic signals and ½FWHM from six MNP samples

| Vial # | I (15 μM) | II (7.5 μM) | III (3.8 μM) | IV (1.5 μM) | V (75 nM) | VI (0 nM) |
|---|---|---|---|---|---|---|
| $f_{H,crit}$ at peak harmonic amplitude (Hz) | 2000 | 2500 | 2700 | 2800 | 2850 | 2900 |
| ½FWHM* (Hz) | 11,300 | 11,800 | 12,300 | 12,700 | 12,900 | 13,100 |

*½FWHM is the difference between $f_{H,crit}$ at peak harmonic amplitude and $f_{H,50\%}$ at 50% of the peak harmonic amplitude.

On the other hand, the ratios of the 3rd to the 5th harmonics (R35) at different drive field frequencies have also been used as a MNP quantity-independent metric for characterizing the biomarker concentrations/quantities from samples (the harmonic ratio model can be found in Note S6 in the Supporting Information). Figure 3(a) summarizes the harmonic ratios, R35, from six samples as we vary the drive field frequency $f_H$ from 500 Hz to 20 kHz. The harmonic ratios R35 increases as the concentrations/quantities of streptavidin increases. The inset figure in Figure 3(a) lists the harmonic ratios R35 from six samples at a drive field frequency of $f_H = 10\ kHz$. The R35 drops from 1.7978 for sample I (streptavidin concentrations: 15 µM) to 1.4332 for sample VI (streptavidin concentrations: 0 nM). Figure 3(b) – (e) show the relationships between R35 and streptavidin concentrations under different drive field frequencies. It is noted that the harmonic ratio R35 increases as the concentrations/quantities of streptavidin increases in the MNP sample. Under low drive field frequency, $f_H = 500\ Hz$ (Figure 3(b)), the difference in R35 between sample I (streptavidin concentrations: 15 µM) and VI (streptavidin concentrations: 0 nM) is only 6%. On the other hand, under high drive field frequency, $f_H = 18\ kHz$ (Figure 3(e)), this difference reaches to 28%. Hence, the harmonic ratio, R35, under higher drive field frequency allows for better resolutions in characterizing the target biomarker concentrations.

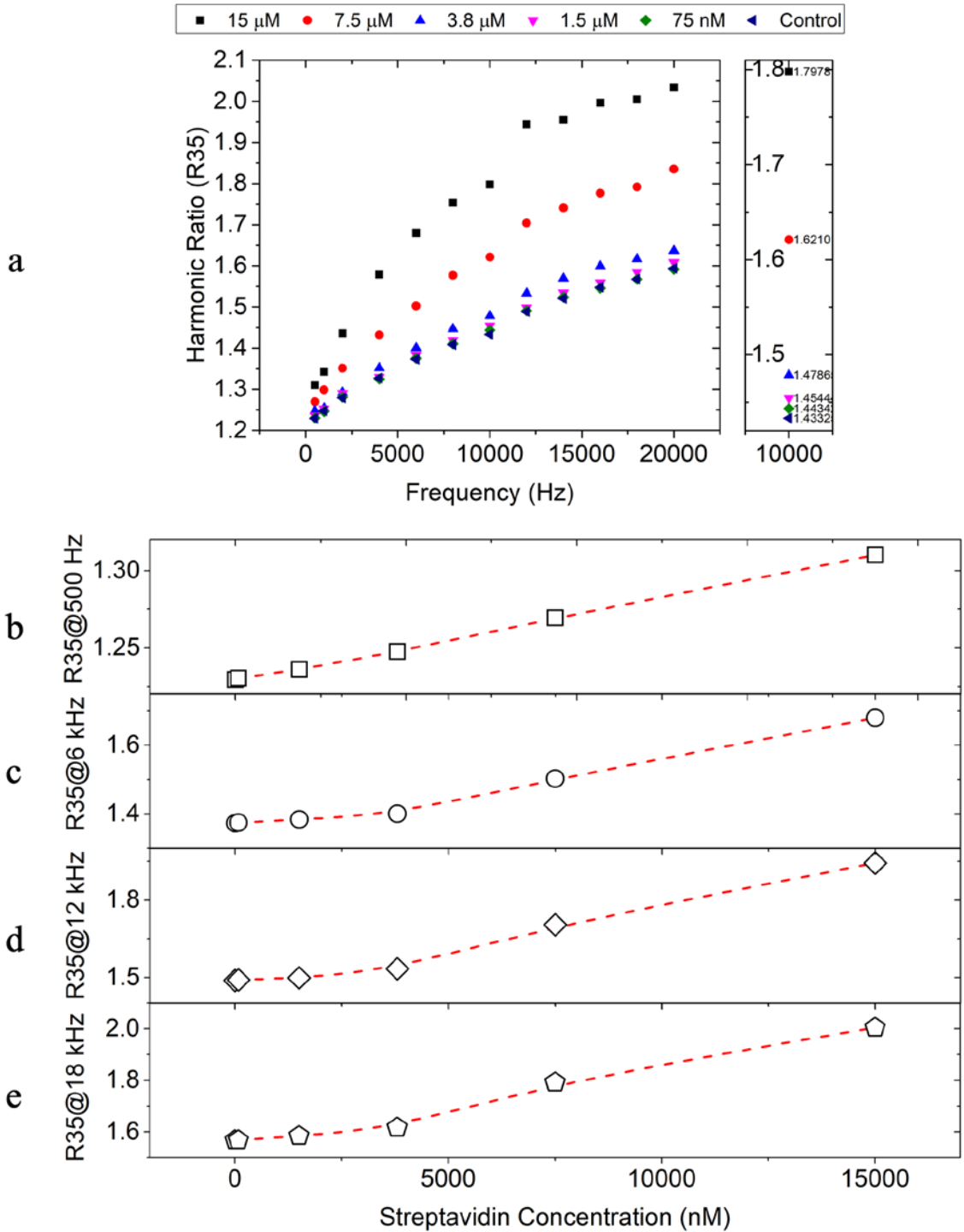

Figure 3. MPS measurements of the ratio of the 3rd to the 5th harmonics (R35) of six MNP samples in different bound states. (a) The harmonic ratios, R35, from six samples as we vary the $f_H$ from 500 Hz to 20 kHz. The inset figure summarizes the R35 under a drive field frequency of $f_H = 10\ kHz$ for vials I – VI, the R35 values are: 1.7978 (15 μM), 1.6210 (7.5 μM), 1.4786 (3.8 μM), 1.4544 (1.5 μM), 1.4434 (75 nM), and 1.4332 (control, 0

nM), respectively. (b) – (d) show the relationships between R35 and streptavidin concentrations at drive field frequencies of (b) 500 Hz; (c) 6 kHz; (d) 12 kHz; and (e) 18 kHz.

Besides the peak shift, ½FWHM, and harmonic ratio (R35) methods, the harmonic angle is also a MNP concentration/quantity-independent metric for characterizing the bound states of MNPs in MPS.[46] The 3$^{rd}$ and the 5$^{th}$ harmonic angles are reconstructed from the total signals and background noise, they are summarized in Note S7 in the Supporting Information.

**Morphological Characterization of MNPs in Different Bound States.** Transmission electron microscopy (TEM) images are taken to investigate the MNP clusters and bound states in six samples after the MPS measurements. A droplet of the MNP solution (~ 10 μL) is dipped onto a TEM grid (copper mesh with amorphous carbon film) with filter paper underneath. The MNP droplet forms a thin liquid layer on the TEM grid and will be ready for TEM characterization when the solution evaporates. Then the TEM grids are characterized in TEM (FEI Tecnai T12, 120 kV). Well-dispersed MNPs are observed from the control sample without streptavidin (vial VI, 0 nM) as seen in Figure 4 (i) & (k) while more MNP clusters could be found with the increase of streptavidin concentrations. As shown in Figure 4 (e) - (i), increased streptavidin concentration produces larger MNP clusters as observed in TEM images. Figure 4 (a) shows different bound states of MNPs in the presence of streptavidin. Since there are multiple biotins on each MNP and the tetrameric structure of streptavidin hosting 4 biotin bindings sites, MNPs could form clusters, chains, tetramers, trimers, dimers, etc. Figure 4 (b) - (d) show the corresponding models of MNP bound states from Figure 4 (a). Figure 4 (j) & (k) are the zoomed in views of one "כ-shape" MNP cluster from vial I and the well-dispersed MNPs from vial VI.

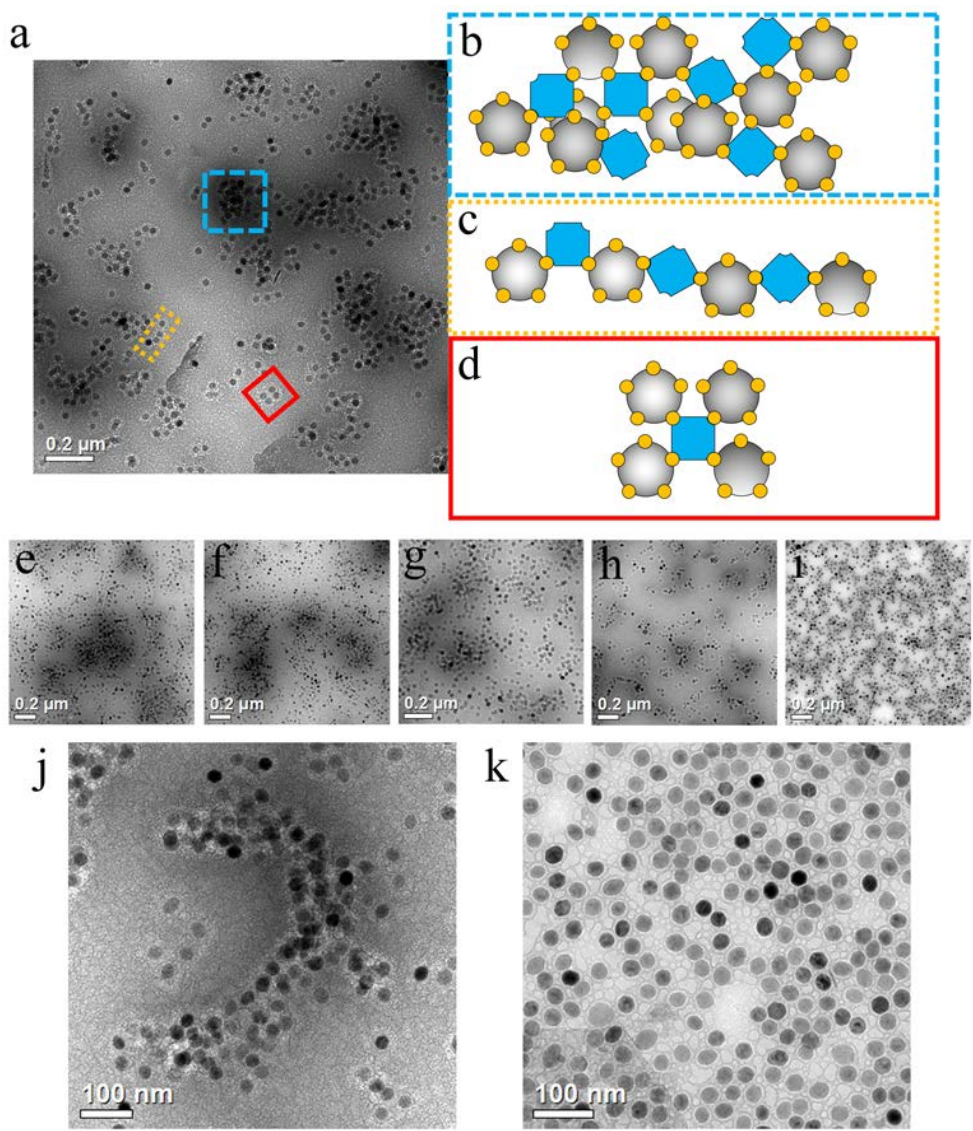

Figure 4. Bright-field TEM images of six MNP samples in different bound states. (a) the sample I (15 μM) showing different MNP bound states. (b) MNP cluster model representing the blue dashed contour in (a). (c) MNP chain model representing the orange dotted contour in (a). (d) MNP tetramer model representing the red solid contour in (a). Blue square represents streptavidin, orange dot represents biotin, and grey circle represents MNP. (e-i) samples I (15 μM), II (7.5 μM), III (3.8 μM), V (75 nM), and VI (0 nM) showing the cluster size increases with the number of streptavidin increases. (j, k) comparison of sample I (15 μM) and VI (0 nM) showing one of the MNP clusters in the presence of streptavidin and the well dispersed MNPs in the absence of streptavidin.

**Hydrodynamic Size Analysis of MNPs in Different Bound States.** Dynamic light scattering (DLS) measurements confirmed the streptavidin-specific clustering and the hydrodynamic size increased with streptavidin concentration. Statistic results from Figure 5 give us the mean hydrodynamic sizes of vials I - VI: 706, 102.8, 63.2, 58.8, 58.4, and 58.3 nm, respectively. The results from DLS are in good agreement with our MPS and TEM measurements.

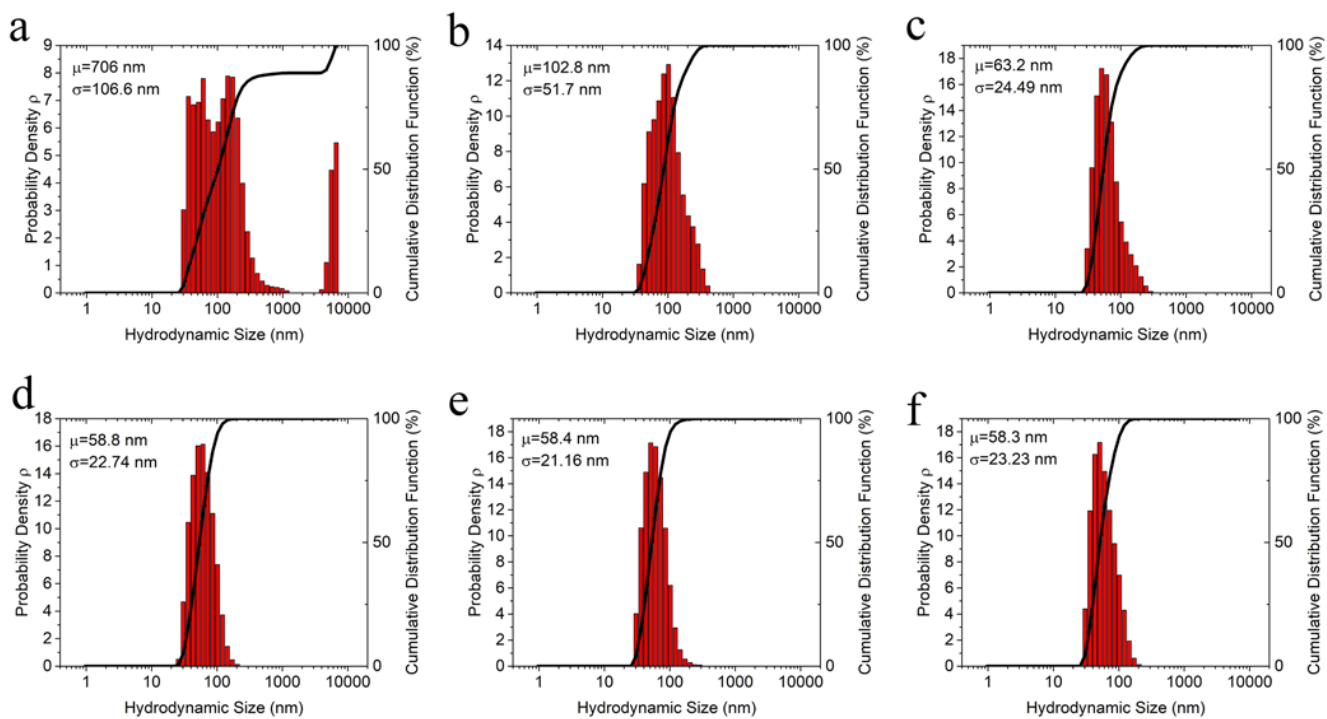

Figure 5. Statistical hydrodynamic size distribution collected using DLS. (a) - (f) are vials I - VI. Solid black lines are cumulative distribution curves, μ and σ denote the mean and the standard deviation of hydrodynamic sizes.

4. **Conclusions and Future Perspectives**

In this work, we have demonstrated the feasibility of using MNP relaxation dynamics-based MPS method for bioassay applications. The specific binding of target analytes onto MNP surface inhibits their rotational freedom, thus, changes the MPS pattern. The streptavidin and biotin system are applied in this study as a model system. Each streptavidin has binding sites for multiple MNPs and forms MNP clusters that tremendously increase the hydrodynamic sizes of MNPs, as a result, noticeable changes could be found in the harmonic phase lags and amplitudes. Herein, we have reported three MNP concentration/quantity-independent metrics for characterizing the bound states of MNPs in MPS, namely, the $f_{H,crit}$ at peak harmonic amplitude (unit: Hz), ½FWHM (unit: Hz), and the ratios of the 3$^{rd}$ to the 5$^{th}$ harmonics (R35). The $f_{H,crit}$ at peak harmonic amplitude and ½FWHM metrics require the full screening of drive field frequency $f_H$ from several hundred Hz to several tens' kHz (in this work we varied $f_H$ from 500 Hz to 20 kHz). Although it is time-consuming to collect the MPS responses under varying drive field frequencies, $f_{H,crit}$ and ½FWHM metrics can provide high sensitivity bioassays if $f_H$ is varied with a small step width (i.e., better frequency resolution). On the other hand, the harmonic ratio R35 provides an alternative way to carry out faster MPS bioassay measurements since it does not require the full screening of the drive field frequencies. As shown in Figure 3(b) – (e), it is possible to collect the R35 under one drive field frequency and use this metric to characterize the target biomarker concentration/quantity. However, the harmonic ratio R35 is unable to provide very high sensitivity bioassays compared to $f_{H,crit}$ and ½FWHM metrics due to the factor that R35 from MNP samples with very low biomarker concentrations are very close to the control sample and it is susceptible to the system noise from MPS and the environmental noise. Although these subtle differences in the R35 from MNP samples with very low biomarker concentrations might be ruined by the noise, it's possible to avoid this issue by using a higher accuracy MPS system with lower system noise and carrying out the measurements in a magnetic shielding room. In addition, it is worthwhile to mention that all the three MNP concentration/quantity-independent metrics we reported in this work are temperature-dependent. Thus, the MPS measurements should be carried out under the same testing temperature to reduce the variances between different MPS-based bioassays.

The MNP concentration/quantity-independent metrics used in this work achieved a sensitivity of 75 nM for the detection of streptavidin using our MPS system (7.5 pmole streptavidin). There are several advantages in this MPS method: wash-free and easy-to-use bioassay (measurements could be carried out on minimally processed biological samples by non-technicians with minimum training requirements), rapid (the testing time is within 10 seconds), cheap (the cost for each trail only requires microgram quantities of iron oxide nanoparticles), and portable (the coils, amplifiers, filters, and DAQ could be assembled onto a single PCB board).

Future development in this MNP relaxation dynamics-based bioassay method is to improve the sensitivity of MPS measurements as well as increase the signal-to-noise ratio (SNR). In summary, this MNP relaxation dynamics-based MPS method for bioassay applications opens a door for future point-of-care, versatile, sensitive, rapid, and wash-free molecular sensing.

**ASSOCIATED CONTENT**

Supporting Information

Note S1. Langevin Model of Magnetic Response

Note S2. Néel and Brownian Relaxation Models

Note S3. Phase Lag Model

Note S4. Phasor Theory

Note S5. Induced Signal Model

Note S6. Harmonic Ratio Model

Note S7. MPS Measurements of Harmonic Angles

Note S8. Recorded Real-time Magnetic Response from Pick-up Coils

Note S9. The Normalized $3^{rd}$ and $5^{th}$ Harmonics as a MNP Quantity-independent Metric for Biomarker Detection

**AUTHOR INFORMATION**

**Corresponding Authors**

*E-mail: wuxx0803@umn.edu (K. W.)

*E-mail: jpwang@umn.edu (J.-P. W.)

**ORCID**

Kai Wu: 0000-0002-9444-6112

Jinming Liu: 0000-0002-4313-5816

Diqing Su: 0000-0002-5790-8744

Renata Saha: 0000-0002-0389-0083

Jian-Ping Wang: 0000-0003-2815-6624

**Notes**

The authors declare no competing financial interest.

**ACKNOWLEDGMENTS**

Portions of this work were conducted in the Minnesota Nano Center, which is supported by the National Science Foundation through the National Nano Coordinated Infrastructure Network (NNCI) under Award Number ECCS-1542202. Portions of this work were carried out in the Characterization Facility, University of Minnesota, a member of the NSF-funded Materials Research Facilities Network (www.mrfn.org) via the MRSEC program.

TOC

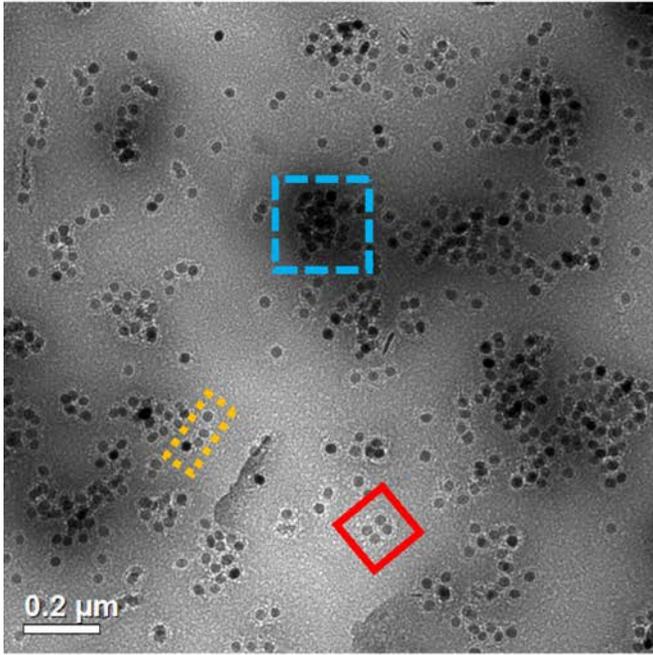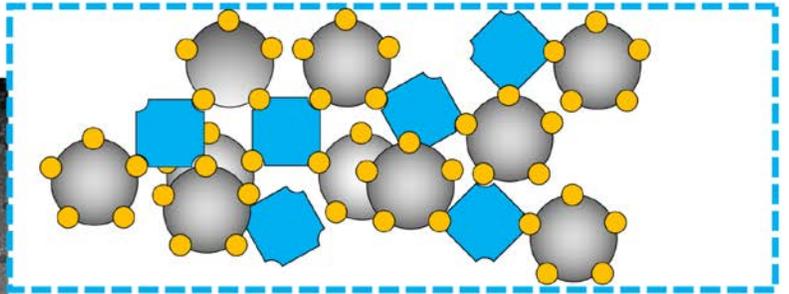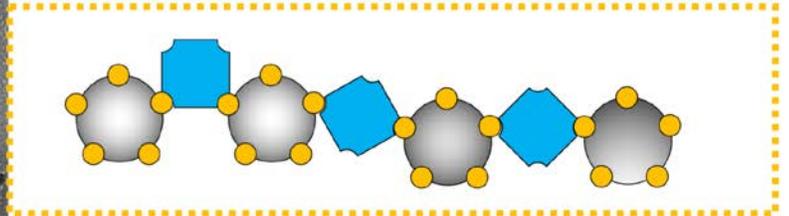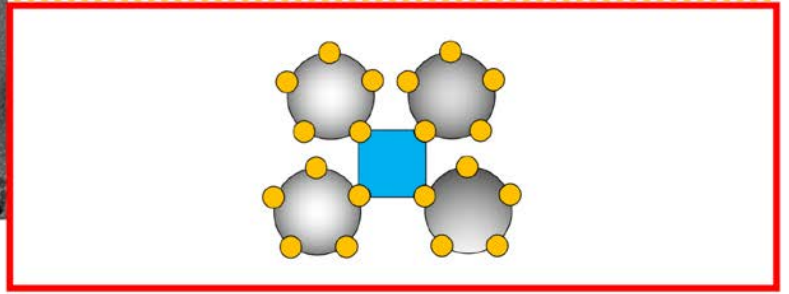